\documentclass{article}
\usepackage[utf8]{inputenc}
\usepackage{authblk}
\usepackage[dvips]{graphicx}
\graphicspath{{noiseimages/}}
\usepackage{xcolor}
\usepackage[T2A]{fontenc}
\usepackage{tikz}
\usepackage{pgfplots}
\pgfplotsset{compat=1.7}
\usetikzlibrary{intersections, pgfplots.fillbetween}
\usetikzlibrary{decorations.pathmorphing}
\usepackage[mode=buildnew]{standalone}
\usepackage[toc,page]{appendix}
\usepackage{amsmath}
\usepackage{amssymb}
\usepackage{float}
\usepackage{comment}
\usepackage{a4wide}
\usepackage[english]{babel}
\usepackage{hyperref}
\definecolor{urlcolor}{HTML}{990000}
\definecolor{linkcolor}{HTML}{005F5F} 
\hypersetup{pdfstartview=FitH,  linkcolor=linkcolor,urlcolor=urlcolor, colorlinks=true,citecolor=blue}
\usepackage[page]{appendix}

\renewcommand{\hat}{}

\author[1,2]{E.T.Akhmedov}
\affil[1]{Moscow Institute of Physics and Technology, Institutskii per. 9, 141700, Dolgoprudny, Russia}
\affil[2]{ Institute for Theoretical and Experimental Physics, B. Cheremushkinskaya 25, 117218, Moscow, Russia}

\title{Curved space equilibration vs. flat space thermalization \\ (a short review)}

\begin{document}

\maketitle

\begin{abstract}

We discuss equilibration process in expanding universes as compared to the thermalization process in Minkowski space--time. The final goal is to answer the following question: Is the equilibrium reached before the rapid expansion stops and quantum effects have a negligible effect on the background geometry  or stress--energy fluxes in a highly curved early Universe have strong effects on the expansion rate and the equilibrium is reached only after the drastic decrease of the space--time curvature? We argue that consideration of more generic non--invariant states in theories with invariant actions is a necessary ingredient to understand quantum field dynamics in strongly curved backgrounds. We are talking about such states in which correlation functions are not functions of such isometry invariants as geodesic distances, while having correct UV behaviour. The reason to consider such states is the presence of IR secular memory effects for generic time dependent backgrounds, which are totally absent in equilibrium. These effects strongly affect the destiny of observables in highly curved space--times.

\end{abstract}




{\bf 1.} To date, quantum field theory (QFT) is the best tool we have to describe the world of fundamental physics. One of the practical goals of QFT is to calculate correlation functions (CF). CF of elementary fields are building blocks of observables. E.g. using the external leg amputation procedure one can find scattering amplitudes from CF in high energy particle physics. In condensed matter theory, in turn, using CF one can calculate stress--energy fluxes or densities of electric currents. 

In high energy particle physics one uses only Poincar\'e invariant states, for which CF are analytic functions of Lorentz invariant variables, which consist of geodesic distances between their arguments and signs of time differences:

\begin{eqnarray}
\Big\langle \phi(\underline{x}_1) \dots \phi(\underline{x}_n) \Big\rangle_0 = F_0\left(\Big\{\left|\Delta\underline{x}_{jk}\right| \pm i \, 0 \, \text{sign} \Delta t_{jk}\Big\}\right), 
\label{PoincInv} \\
{\rm where} \quad \underline{x} = (t, \, \vec{x}) \quad \text{and} \quad \Delta\underline{x}_{jk} = \underline{x}_j - \underline{x}_k, \quad j = \underline{1,n}. \nonumber 
\end{eqnarray}
In condensed matter theory one frequently restricts attention to stationary (thermal) states. CF over such states depend only on the time differences rather than on each their time argument separately:

\begin{equation}
\Big\langle \phi(\underline{x}_1) \dots \phi(\underline{x}_n) \Big\rangle_T = F_T\Big(\Big\{\Delta t_{jk}\Big\}\Big| \vec{x}_1, \dots, \vec{x}_n\Big), \quad \Delta t_{jk} = t_j - t_k.\label{therm}
\end{equation}
If the state is not spatially homogeneous, then such CF can depend on each their spatial argument separately, as we have explicitly shown here. {\bf Note that the action of the theory can be even Poincar\'e invariant, but one can consider a state in such a theory that does not respect the symmetry:} e.g. the thermal state, in which CF being invariant under time translations are not Poincar\'e invariant. Meanwhile CF (\ref{therm}) over such states have proper (Hadamard) behavior for the light--like separations of their points. The UV behaviour of CF in any background state should be the same as in Minkowski space--time for Poincar\'e invariant state, if the action of the theory is invariant. After all, very high frequency quanta should not sense neither background state nor the background geometry.

Actually one usually considers states which are very close to homogeneity, when CF are slow functions of $\left(\vec{x}_j + \vec{x}_k\right)/2$ and rapid ones of $\Delta \vec{x}_{jk}$ \cite{LL10}. Slightly perturbed thermal CF out of equilibrium possess the same property in time variables, $t_1, \dots, t_n$. Namely they depend slowly on $(t_j+t_k)/2$ and rapidly on $\Delta t_{jk}$. (This situation is referred to as the kinetic approximation.) At future infinity CF relax to the form described by (\ref{therm}).

Furthermore, the common wisdom is to consider isometry invariant states \cite{Mottola:1984ar}, \cite{Marolf:2010zp}, \cite{Marolf:2010nz}, \cite{Higuchi:2010xt}, \cite{Moreau:2018lmz}, \cite{Guilleux:2015pma}, \cite{Hollands:2010pr} or slight deviations from them \cite{Starobinsky:1994bd}, \cite{Krotov:2010ma}, \cite{Tsamis:2005hd}, \cite{Gorbenko:2019rza} in de Sitter (dS) space--time. In such states the situation is similar to (\ref{PoincInv}): CF either are analytic functions of the dS isomerty invariants or reach such a stage at future infinity. Here and throughout this note we are talking about the situation when gravity can be treated semiclassically, while other fields should be quantized.

There is a good reason for such a point of view based on our everyday experience: present day Universe has a very small curvature and we mostly deal with processes which are very close to equilibrium, even if they are nonstationary. At least we see only such processes which quickly lead to thermalization without any strong backreaction on weak background fields, observed in the Universe at this stage of its evolution. 

{\bf Meanwhile in the early Universe, if its metric was truly very close to dS one with a GUT scale curvature, the situation was highly nonstationary.}  There is no reason that the rapid expansion has started from a highly symmetric state with isometry invariant CF. The presence of galaxies and other inhomogeneities seen in the present day Universe are a direct sign that the symmetry was violated somehow at some stage, even if the background gravitational field was exactly a highly symmetric one. It goes without saying that obviously the background metric was only approximately close to the dS one.

Thus, there is no reason that the corresponding CF in early Universe to have been isometric invariant or even close to that. In generic situations CF should have been functions of each their argument separately and were dramatically changing in space and time during the course of rapid expansion\footnote{Actually the same should be true in the vicinity of microscopic (primordial) black holes \cite{Akhmedov:2015xwa} or in strong background fields of any other nature \cite{Akhmedov:2014hfa}, \cite{Akhmedov:2014doa}, \cite{Akhmedov:2020haq}.}:  

\begin{equation}
\Big\langle \phi(\underline{x}_1) \dots \phi(\underline{x}_n) \Big\rangle_{U} = F_U\Big(\underline{x}_1, \dots, \underline{x}_n\Big). \label{gener}
\end{equation}
This situation can appear in theories with invariant actions and with proper UV behaviour of CF for light--like separations of their points. 

Note that at late stages of rapid expansion such CF can become almost spatially homogeneous. However, there is no reason for CF to become simultaneously stationary, when the metric is rapidly changing in time:

\begin{equation}
\Big\langle \phi(\underline{x}_1) \dots \phi(\underline{x}_n) \Big\rangle_{HU} \approx F_{HU}\Big(t_1, \dots, t_n\Big|\Big. \left\{\Delta\vec{x}_{jk}\right\} \Big).
\end{equation}
{\bf When the curvature of our Universe was of GUT scale, observables must have been various stress-energy fluxes rather than scattering cross--sections.} Such observables can be measured only indirectly --- via their backreaction on the background geometry or other fields. 

Pay attention that we do not assume that stress--energy fluxes in early Universe were separable into anything like particles. In fact, in those circumstances any free Hamiltonian was a fast function of time and was not diagonal in any region of space--time. Thus, there was no any reason even for the composition principle to work. 

{\bf 2.} QFT does provide a tool to calculate CF even in such unusual conditions. This tool does not assume that background field is adiabatically turned on. Neither it assumes that the background field is weak enough to provide a very small particle creation. The method does work without such approximations, which are probably meaningless in early Universe or at least in very strong background fields.

The difference with respect to the equilibrium cases is that to get an unambiguous answer for CF in non--stationary situations one has to specify an initial Cauchy surface, a basis of modes, an initial state (prepared with the use of the creation and annihilation operators corresponding to the modes) and then use the Schwinger-Keldysh rather than the Feynman diagrammatic technique. The above ingredients provide the causal evolution of CF with the change of each of their argument as solutions of the initial value problem. In a sense, in such a situation there is a clear direction of time with a unitary evolution.

{\bf In fact, in the described circumstances loop corrections do not just lead to UV renormalization of various coupling constants and masses, they also contain IR secular memory effects, which are totally absent in equilibrium.} These memory effects are sensitive to the choice of a patch of the entire space--time and on the initial (and boundary) data. That is a very unusual phenomenon as compared to the equilibrium cases.

Thus, fair questions one can ask now are as follows: what was the state at the very beginning of the rapid expansion of the Universe? Was equilibrium reached before the end of expansion and without any strong backreaction on the geometry? Or may be stress--energy fluxes did have a strong effect on the expansion rate and the equilibrium was reached only after the drastic decrease of the background curvature? More generally: is the backreacion on the geometry negligible for any initial state and expanding background or there are such states, obeying reasonable (e.g. Hadamard) conditions, for which the backreaction is not negligible? 
Or even may be the backreaction in a generic highly curved expanding space--time is not negligible for generic initial state?

Why these questions are important for the physics in the early Universe? The point is that, if CF are not isometry invariant, then expectation values of stress--energy tensors of various fields, $\left\langle T_{\mu\nu} \right\rangle$, are not proportional to the metric tensor, $g_{\mu\nu}$. Furthermore, secular memory effects make quantum loop corrections for CF to become of the same order as classical tree--level contributions. These contributions can strongly affect such expectation values as $\left\langle T_{\mu\nu} \right\rangle$ and can lead to the violation of the Ehrenfest theorems. In such a case there can be non--trivial fluxes, which can lead to a screening of the cosmological constant independently of its physical origin.

As we argue in this note, intuition gained by considerations of only stationary or almost stationary examples in Minkowski space--time is not of any help in answering these questions. {\bf To show our point in this note we consider several examples gradually increasing their difficulty.}

{\bf 3.} To be a bit more specific let us consider a situation in which one has to find the time evolution of the expectation value of an operator. By definition the time evolution of a CF is:

\begin{equation}
\left\langle \hat{O}\right\rangle(t) \equiv \left\langle \psi_0\left| \, \overline{T} e^{i \, \int_{t_0}^t dt' \, \hat{H}(t')} \, \hat{O} \, T e^{- i \, \int_{t_0}^t dt' \, \hat{H}(t')} \, \right| \psi_0 \right\rangle.\label{main}   
\end{equation}
In this equation $\hat{H}(t) = \hat{H}_0(t) + \hat{V}(t)$ is the full Hamiltonian of the theory. The average of the operator over the initial state $\left| \psi_0 \right\rangle$ is given as the initial value for the problem: $\left\langle \hat{O}\right\rangle(t_0) = \hat{O}_0.$

Here for simplicity we consider a single operator $\hat{O}$, ignore its spatial dependence and take the average over a pure initial state. The extension of our discussion to multi--point CF and to mixed states is straightforward.

Changing to the interaction representation, eq. (\ref{main}) gets converted into:

\begin{equation}
\left\langle \hat{O}\right\rangle(t) = \Big\langle \psi_0\Big| \, \hat{S}^+(t, t_0) \, \hat{O}_0(t) \, \hat{S}(t, t_0) \, \Big| \psi_0 \Big\rangle, \quad \text{where} \quad \hat{S}(t,t_0) = T e^{ - i \, \int_{t_0}^t dt' \, \hat{V}_0(t')},    
\end{equation}
and $\hat{O}_0(t)$, $\hat{V}_0(t)$ are $\hat{O}$, $\hat{V}$ operators taken in the interaction picture.

At this stage in the last equation one usually inserts the unit operator of the form $1 = \hat{S}^+(+\infty, t) \, \hat{S}(+\infty, t)$ between $\hat{S}^+(t,t_0)$ and $\hat{O}_0(t)$ to obtain:

\begin{equation}
\left\langle \hat{O}\right\rangle(t) = \Big\langle \psi_0\Big| \, \hat{S}^+(+\infty, t_0) \, T\Big[\hat{O}_0(t) \, \hat{S}(+\infty, t_0)\Big] \, \Big| \psi_0 \Big\rangle. \label{main1}
\end{equation}
The dependence on $t_0$ is of crucial importance here. To see our point let us consider, first, the well known situation when the dependence on $t_0$ does disappear from (\ref{main1}). Here are the conditions for this to happen:

\begin{enumerate}
    \item The normal ordered free Hamiltonian $\hat{H}_0$ is time independent and bounded from below; 
    \item The expectation value in (\ref{main1}) should be taken over the ground state of $\hat{H}_0$: $\left| \psi_0 \right\rangle = \left| 0 \right\rangle$, $\hat{H}_0 \, \left| 0 \right\rangle = 0$;
    \item Interaction term, $\hat{V}$, is turned on adiabatically after $t_0$. Then $\hat{V}$ also should be switched off adiabatically after $t$. 
\end{enumerate}

In effect we have to make in (\ref{main1}) the substitution as follows: $\hat{S}(+\infty, t_0) \to \hat{S}_{tt_0}(+\infty, -\infty)$. Before such a substitution the evolution was considered as starting abruptly right after $t_0$.

If all these conditions are fulfilled, the Fock space ground state $\left| 0 \right\rangle$ remains intact under the action of the evolution operator $\hat{S}_{tt_0}(+\infty, -\infty)$:

\begin{equation}
\left|\left\langle 0 \left| \hat{S} \right| 0 \right\rangle\right| = 1 \quad \text{and} \quad \left\langle n \neq 0 \left| \hat{S} \right| 0 \right\rangle = 0, \quad \text{where} \quad \hat{S} \equiv \hat{S}_{tt_0}(+\infty, -\infty),  \label{cond}
\end{equation}
and $\left\langle n \right|$ is the basis of eigen--states of $\hat{H}_0$. 

This way the dependence of the CF on $t_0$ does disappear at every loop order. In fact, then one can insert into (\ref{main1}), between $S^+$ and the $T$--ordered product, the unit operator of the form $1 = \sum_n \left| n \right\rangle \, \left\langle n \right|$ to obtain:

\begin{eqnarray}
\left\langle \hat{O}\right\rangle(t) = \sum_n \, \Big\langle 0\Big| \, \hat{S}^+ \, \Big| n \Big\rangle \, \Big\langle n \Big| \, T\Big[\hat{O}_0(t) \, \hat{S}\Big] \, \Big| 0 \Big\rangle = \frac{\Big\langle 0 \Big| \, T\Big[\hat{O}_0(t) \, \hat{S}\Big] \, \Big| 0 \Big\rangle }{\Big\langle 0\Big| \, \hat{S} \, \Big| 0 \Big\rangle}.\label{Feynexp}
\end{eqnarray}
Here at the last step we have used (\ref{cond}). Note that in such a situation all CF do not contain any anti--time--ordered expressions. Only in this very special (stationary) situation one can apply the Feynman diagrammatic technique to calculate such CF as (\ref{PoincInv}). 

Another stationary option, in which the dependence on $t_0$ does disappear from CF at every loop order, is when instead of (\ref{main1}) one takes the average over the plankian density matrix. Below we come back to consideration of such a case on a concrete example. Furthermore, below we also will describe the third stationary option which is specific to dS space--time.

If any of the listed above conditions 1-3 is violated, one has to deal directly with (\ref{main1}) and to perturbatively expand both T--ordered $\hat{S}$ and anti--T--ordered $\hat{S}^+$. In such a case in loop corrections there is the sum over Feynman diagrams, but with Ising model type of ``$\mp$'' signs attached to every vertex in them. The sign depends on whether the vertex comes from $\hat{S}$ or $\hat{S}^+$. 

This is how one arrives at the Schwinger--Keldysh diagrammatic technique \cite{LL10}, \cite{Kamenev}. As one can see, in such a case contributions to CF are strictly causal, depend on the position of the initial Cauchy surface, $t_0$, and do not demand any adiabatic treatment of the interaction term.

{\bf 4.} To be even more specific in illustrating our points, but to keep our presentation simple and quite general, we consider selfinteracting real massive scalar field theory:

\begin{equation}
S = \int d^Dx \, \sqrt{|g|}\, \left[\frac12 \, \partial_\mu \phi\, \partial^\mu \phi + \frac{m^2}{2} \,\phi^2 + \frac{\lambda}{4} \, \phi^4 \right] \label{action}
\end{equation}
in a $D$--dimensional space--time with spatially homogeneous metric: 

\begin{equation}
    ds^2 =  - dt^2 + a^2(t) \, d\vec{x}^2. \label{metric}
\end{equation}
We will only vaguely mention spatially inhomogeneous situations throughout the review. For our discussion the UV renormalizability of the theory is not important. We will consider only such entities of CF which are UV finite. We assume that mass, $m$, and selfcoupling, $\lambda$, acquire their UV renormalized (physical) values. 

The mode expansion of the field operator is as follows:

\begin{equation}
\hat{\phi}(t,\vec{x}) = \int \frac{d^{D-1}\vec{p}}{(2\, \pi)^{D-1}} \, \Big[\hat{a}_{\vec{p}} \, f_p(t) \, e^{- i \, \vec{p} \, \vec{x}} + \text{h.c.}\Big]. \label{phi}
\end{equation}
Here the creation and annihilation operators, $\hat{a}_{\vec{p}}$ and $\hat{a}_{\vec{p}}^+$, obey the Heisenberg algebra if $\hat{\phi}$ obeys the canonical commutation relations with the corresponding conjugate momentum and the modes, $f_p(t)\, e^{- i \, \vec{p} \, \vec{x}}$, solve the Klein--Gordon equation in the metric under consideration.

There are, actually, other ways to perform quantization: e.g. one can consider a basis of spatial modes on the Cauchy surface at $t_0$ and demand that corresponding $\hat{a}_{\vec{p}}$ and $\hat{a}_{\vec{p}}^+$ obey the Heisenberg algebra; and only then one can evolve the field operator $\hat{\phi}$ and the corresponding conjugate momentum with the free Hamiltonian. Then the time dependence inside the mode expansion of $\hat{\phi}(t,\vec{x})$ can be more complicated than in (\ref{phi}). 

In flat Minkowski space--time, when $a(t) = 1$ in (\ref{metric}), both ways to quantize the theory lead to the same answer. But that is not necessarily true in generic backgrounds for generic initial basis of modes on the initial Cauchy surface. To respect the isometry of the background space--time it is necessary to adopt the way presented in the paragraph around eq. (\ref{phi}). But if there is no any symmetry, then any of the two ways is equally applicable. It is an open question which of them one has to use in the settings under consideration. 

Let the initial state is spatially homogeneous and with such a density matrix $\hat{\rho}^0$ that:

\begin{eqnarray}
\left\langle\hat{a}_{\vec{p}}^+ \, \hat{a}_{\vec{q}} \right\rangle \equiv \sum_n \rho^0_n \, \left\langle n\left| \hat{a}_{\vec{p}}^+ \, \hat{a}_{\vec{q}} \right| n \right\rangle = n^0_p\, \delta\left(\vec{p} - \vec{q}\right), \quad \left\langle\hat{a}_{\vec{p}}^+ \, \hat{a}_{\vec{q}}^+ \right\rangle^* = \Big\langle\hat{a}_{\vec{p}} \, \hat{a}_{\vec{q}} \Big\rangle = \kappa^0_p\, \delta\left(\vec{p} + \vec{q}\right), \label{initial}
\end{eqnarray}
where $\hat{H}_0(t_0) \, \left| n \right\rangle = E_n \, \left| n \right\rangle$ and $n^0_p$ is the level--population for the exact modes, which is not necessarily equal to the plankian one, while $\kappa_p^0$ is an anomalous quantum average\footnote{Actually, technically the extension of our considerations to spatially inhomogenous metrics and states is straightforward and not very hard, if the Wick theorem still works. In such a case one has $\left\langle\hat{a}_{\vec{p}}^+ \, \hat{a}_{\vec{q}} \right\rangle = n_{\vec{p}, \vec{q}}$ and $\Big\langle\hat{a}_{\vec{p}} \, \hat{a}_{\vec{q}} \Big\rangle = \kappa_{\vec{p}, \vec{q}}$. (See e.g. \cite{Radovskaya:2020lns} on how to deal with generic initial states.) However, scientific community does not yet have much of intuition to interpret the obtained results in such a case, due to the lack of solved examples.}. Let us stress that in this note we always consider so called Hadamard states for which $n_p^0, \kappa_p^0 \to 0$ as $\left|\vec{p}\right|\to \infty$. CF for such states have proper (flat space) UV behaviour for light--like separation between their points.

In Schwinger--Keldysh technique every field is characterized by a matrix of propagators \cite{LL10}, \cite{Kamenev}. After the spatial Fourier transformation the entities of the matrix are as follows:

\begin{eqnarray}
D^{--}\Big(t_1, \, t_2\Big|\Big. \vec{p}\Big) = i \, \left\langle T \Big(\hat{\phi} (t_1, \vec{p}) \, \hat{\phi}(t_2, - \vec{p})\Big) \right\rangle, \quad D^{++}\Big(t_1, \, t_2\Big|\Big. \vec{p}\Big) = i \, \left\langle \overline{T} \Big(\hat{\phi}(t_1, \vec{p}) \, \hat{\phi}(t_2, - \vec{p})\Big) \right\rangle, \nonumber \\
D^{-+}\Big(t_1, \, t_2\Big|\Big. \vec{p}\Big) = i \, \Big\langle \hat{\phi}(t_1, \vec{p}) \, \hat{\phi}(t_2, - \vec{p}) \Big\rangle, \quad D^{+-}\Big(t_1, \, t_2\Big|\Big. \vec{p}\Big) = i\, \Big\langle  \hat{\phi}(t_1, \vec{p}) \, \hat{\phi}(t_2, - \vec{p})\Big\rangle.\label{propaga}
\end{eqnarray}
Here in $D^{--}$ both $\phi$'s come from $S$, while in $D^{++}$ --- from $S^+$. At the same time in $D^{\pm\mp}$ one of the $\phi$'s comes from T--ordered $S$, while another --- from anti--T--ordered $S^+$. Actually all four propagators can be written in the T--ordered form, but along the so called Keldysh time contour --- the one going from $t_0$ to plus infinity, due to the presence of $S$, and then back, due to $S^+$.

The four propagators (\ref{propaga}) are not independent from each other: $D^{--} + D^{++} = D^{-+} + D^{+-}$. After the so called Keldysh rotation \cite{LL10}, \cite{Kamenev} in the space of $\phi$'s coming from $S$ and $S^+$ one arrives at three independent propagators. Their tree--level form in the initial state (\ref{initial}) is as follows: 
The retarded propagator is \cite{vanderMeulen:2007ah}, \cite{Akhmedov:2013vka}:

\begin{equation}
   i\, D^R_0\Big(t_1, \, t_2\Big|\Big. \vec{p}\Big) \equiv i\, \theta(t_2 - t_1) \, \Big[\hat{\phi}(t_1, \vec{p}), \, \hat{\phi}(t_2, - \vec{p})\Big]  = \theta(t_2 - t_1) \, \text{Im}\Big(f_p(t_1) \, f^*_p(t_2)\Big), \label{retar}
\end{equation}
the advanced propagator is fixed as $D^A\Big(t_1, \, t_2\Big|\Big. \vec{p}\Big) = D^R\Big(t_2, \, t_1\Big|\Big. \vec{p}\Big)$ and the Keldysh propagator is \cite{vanderMeulen:2007ah}, \cite{Akhmedov:2013vka}:

\begin{eqnarray}
    i \, D^K_0\Big(t_1, \, t_2\Big|\Big. \vec{p}\Big) \equiv \frac{1}{2} \, \Big\langle \Big\{\hat{\phi}(t_1, \vec{p}), \, \hat{\phi}(t_2, - \vec{p}) \Big\} \Big\rangle = \nonumber \\ = \left(\frac12 + n^0_p\right)\, f_p(t_1) \, f^*_p(t_2) + \kappa_p^0 \, f_p(t_1) \, f_{-p}(t_2) + \text{c.c.} \, . \label{Keldysh}
\end{eqnarray}
Under the sign ``c.c'' we also assume the presence of $\kappa_p^{0*} = \left\langle\hat{a}_{\vec{p}}^+ \, \hat{a}_{\vec{q}}^+ \right\rangle$.

One can see that tree--level retarded propagator is state independent and characterizes only the spectrum of quanta in the theory. That is the common situation, because it contains commutator of the fields. The tree--level Keldysh propagator does depend on $n_p^0$ and $\kappa_p^0$, i.e. on the state. 

Let us see now what happens to these CF if one takes into account loops. Consider first the situation in Minkowski space--time, when $\kappa^0_p = 0$ in (\ref{Keldysh}) and when the modes have the standard form

\begin{equation}
f_p(t) = \frac{e^{-i\, \omega_p \, t}}{\sqrt{2\,\omega_p}}, \quad \text{and} \quad \omega_p = \sqrt{\vec{p}^2 + m^2}. \label{modes}    
\end{equation}
To have a clear interpretation of the loop corrections it is instructive to study their expressions in various limits. 

One of the standard limits is $\left|t_2 - t_1\right| \to \infty$ in the characteristic units. In such a situation there is a secular effect, but it is not of memory type. In fact, in this limit the leading loop corrections are not sensitive to the change of the state in the theory: At leading order, as $\left|t_2 - t_1\right| \to \infty$, both retarded and Keldysh propagators receive secular corrections $\sim \left|t_2 - t_1\right|$, which can be absorbed into a mass renormalization or a complex shift of mass. This limit is not of interest for us.

Secular memory effect, which is sensitive to the change of the state in the theory, reveals itself in another limit, when both points of the propagators are taken to the future infinity: 

\begin{equation}
t \equiv \frac{t_1 + t_2}{2} \gg \left|t_1 - t_2\right|. \label{limit}    
\end{equation}
In such a limit the retarded and advanced propagators do not receive any growing with $t$ corrections. That is due to their causal properties, which remain intact even in the loops \cite{Kamenev}. It is only the Keldysh propagator which receives secular corrections as $t-t_0 \to \infty$ in characteristic units. And the correction shows the change of the initial state of the theory.

The first contribution of the memory type comes from the two loop sunset diagram correction to the Keldysh propagator. Namely, when $t-t_0 \to \infty$ the corrected Keldysh propagator can be written in the same form as (\ref{Keldysh}) with $\kappa_p^0 = 0$, but with modified $n_p^0$ in the following way \cite{Akhmedov:2013vka}, \cite{Berges:2004yj}:

\begin{equation}
n^{0+2}_p(t) \approx n_p^0 + \lambda^2 \cdot (t - t_0)\cdot \frac{I\left[n^0\right]}{\omega_p}, \label{correc1}
\end{equation}
where

\begin{eqnarray}
I\left[n^0\right] \propto  \int \frac{d^{D-1}\vec{q}_1 \, d^{D-1}\vec{q}_2 \, d^{D-1}\vec{q}_3}{\omega_1 \, \omega_2 \, \omega_3 } \, \delta^{(D)}\Big(\underline{p} + \underline{q}_1 - \underline{q}_2 - \underline{q}_3 \Big) \times \nonumber \\ \times \Big[\left(1 + n^0_p\right) \, \left(1 + n^0_1\right)\, n^0_2 \, n^0_3 -  n^0_p \, n^0_1 \, \left(1 + n^0_2\right) \, \left(1 + n^0_3\right) \Big], \label{correc}
\end{eqnarray}
and $\underline{q}_a = (\omega_a, \vec{q}_a)$, $\omega_a = \sqrt{\vec{q}_a^2 + m^2}$, and $n^0_a \equiv n^0_{q_a}$, $a = \overline{1,3}$. 

{\bf 5.} The obtained expressions (\ref{correc1}), (\ref{correc}) have the following physical interpretation:

\begin{itemize}

\item Tree--level propagators (\ref{retar}) and (\ref{Keldysh}) in flat space--time are functions of only $t_2 - t_1$. That is because of the form of the modes (\ref{modes}). Thus, at tree--level the time translation invariance is respected, because in the Gaussian theory (without selfinteractions, $\lambda = 0$) all the time--dependence of CF is contained in the modes, while $n^0_p$ remains time independent.
Poincar\'e symmetry is broken, however, if $n_p^0 \neq 0$.

\item In the loops time translation invariance is broken, because the loop corrected CF depends separately on $t_1$ and $t_2$. In fact, the product of the modes in (\ref{Keldysh}) still depends only on $t_1 - t_2$, but now also corrected level--population $n_p^{0+2}$ from (\ref{correc1}) depends on $t = (t_1 + t_2)/2$. 

\item Unlike the stationary situation, the parameter $t_0$ cannot be taken to the past infinity. Otherwise loop corrections will be infinite even after UV regularization. This is what one may call as IR catastrophe: We encounter here the IR divergence, which is cut by the parameter $t_0$. Note that the divergence is present for any value of the mass $m$.

\item Let the initial value $n_p^0$ be of the plankian form:

\begin{equation}
n^{T}_p = \frac{1}{e^{\omega_p/T} - 1}, \label{plank}
\end{equation}
for a temperature $T$. Then it is not hard to see that $I\left[n^T\right] = 0$. This happens because of the energy conservation. In such a case the dependence of CF on $t_0$ and $t_1 + t_2$ disappears at every loop order. This is exactly the thermal stationary situation that has been mentioned above. Only in such situations one can make analytical continuation in $\Delta t = t_1 - t_2$ into Euclidian space. Furthermore, obviously for the ground state one also has that $I[0] = 0$. In such a situation one can make the analytical continuation into the Euclidian space in complex plane of geodesic distance.

\item The origin of the secular growth, as $t-t_0\to \infty$, in (\ref{correc}) can be interpreted as follows. The level--population $n_p$ does not change at tree--level --- in the Gaussian approximation, when $\lambda = 0$. However, in the loops, due to the presence of selfinteractions, there are scattering processes which lead to the redistribution of $n_p$'s. 
Note that all of the particles in (\ref{correc}) are on mass--shell despite the fact that we consider a quantum loop correction. That it because in the IR limit $t-t_0 \to \infty$, which is taken in the correction, the internal legs of the loops relax on mass--shell. Finally note that in $I\left(n^0\right)$ there is the integration over all possible incomes and outcomes of the scattering processes, in which one of the particles is on the level $\underline{p}$. 

\item Even if $\lambda$ is small, after a long enough evolution time, $t-t_0 \to \infty$, the correction (\ref{correc}) becomes of the same order as the tree--level contribution $n_p^0$: $\lambda^2 \, (t-t_0) \, I/\omega \sim 1$. Furthermore one can show that higher loops are also not suppressed. Thus, to understand what is going on, one has to resum at least leading corrections from all loops.

\item Unlike the UV renormalization, in this IR case there is a clear grading between contributing diagrams \cite{Kamenev}, \cite{Akhmedov:2013vka}. Namely, consider the limit $\lambda \to 0$ and $t-t_0 \to \infty$ such that $\lambda^2 \, (t-t_0)$ remains finite. Then one can show \cite{Kamenev} that loop corrections to the retarded and advanced propagators are suppressed by higher powers of $\lambda$ and vanish in the limit under consideration. That is essentially because these propagators do not receive corrections at leading order, as we have pointed out above. Similarly vertexes receive corrections suppressed by higher powers of $\lambda$, which are not accompanied by appropriate powers of $t-t_0$ \cite{Akhmedov:2013vka}, if all their external legs are taken to the future infinity together with $t$.

\end{itemize}

As the result, to resum the leading corrections from all loops, i.e. powers of $\lambda^2 \, (t-t_0)$, it is sufficient to solve only the Dyson--Schwinger (DS) equation for the Keldysh propagator. At leading order one can ignore equations for the retarded and advanced propagators and vertexes. Furthermore, in this single DS equation only the Keldysh propagator can be taken in its exact form. The vertexes, retarded and advanced propagators can take their tree--level form. 

The ansatz for the exact Keldysh propagator is obvious --- it should have the form of (\ref{Keldysh}) with exact $n_p(t)$ instead of $n^0_p$ and with $k_p^0 = 0$. In the kinetic approximation, when $n_p(t)$ is a slow function of time in comparison with the modes, the DS equation acquires the form of the quantum Boltzmann equation \cite{LL10} (see also \cite{Kamenev} and \cite{Berges:2004yj}):

\begin{equation}
    \omega_p \, \frac{n_p(t) - n_p^0}{t - t_0} \approx \omega_p \, \frac{dn_p(t)}{dt} \approx \lambda^2 \, I\Big[n(t)\Big],\label{Boltz}
\end{equation}
where the collision integral $\lambda^2 I$ has the same form as in (\ref{correc}). Let us make a side remark that the extension to the slightly spatially inhomogeneous situation in the kinetic approximation is straightforward \cite{LL10}, \cite{Kamenev}, \cite{Berges:2004yj}:

\begin{equation}
    \omega_p \, \partial_t \, n_p\left(t, \, \vec{x}\right) + \vec{p} \, \vec{\partial} \, n_p\left(t, \, \vec{x}\right) \approx \lambda^2 \, I\Big[n\left(t, \, \vec{x}\right)\Big]. \label{Boltz1}
\end{equation}
These Boltzmann's equations describe the process, which can be predicted on general physical grounds. What we see here is the thermalization --- the evolution of the level population $n_p\left(t, \, \vec{x}\right)$ towards the plankian distribution (\ref{plank}) with the value of $T$, which depends on the initial conditions. The process is irreversible due to the H--theorem, which follows from the properties of the equation (\ref{Boltz1}). Note that in the kinetic approximation the dependence on $t_0$ does disappear.


Of course in real situations there are certain complications related to the fact that (\ref{action}) is only an effective theory. As a result (\ref{Boltz1}) describes the thermalization process only qualitatively rather than quantitatively. But to approach a real situation one just has to adjust such phenomenological entities of the collision integral as scattering cross--sections.

{\bf 6.} Consider now loop corrections for a generic choice of modes $f_p(t)$ in (\ref{phi}) and for the initial values $n_p^0 = 0$ and $\kappa_p^0 = 0$ in (\ref{Keldysh}). As in the situation described in the section {\bf 4}, the secular memory contribution may come from the sunset two--loop diagram correction to the Keldysh propagator. In the limit (\ref{limit}) the corrected propagator has the same form as (\ref{Keldysh}), but with the following expressions in place of $n_p^0$ and $\kappa_p^0$ \cite{Akhmedov:2013vka}:

\begin{eqnarray}
n^{(2)}_p(t) \propto \lambda^2 \, \int\limits_{t_0}^t dt_3 \, a^{D-1}(t_3) \int\limits_{t_0}^t dt_4 \, a^{D-1}(t_4) \, \int d^{D-1}\vec{q}_1 \, d^{D-1}\vec{q}_2 \, d^{D-1}\vec{q}_3 \times  \nonumber \\ \times  \delta^{(D-1)}\Big(\vec{p} + \vec{q}_1 + \vec{q}_2 + \vec{q}_3\Big) \, f_p(t_3) \, f^*_p(t_4) \, f_1(t_3) \, f^*_1(t_4) \, f_2(t_3) \, f^*_2(t_4) \, f_3(t_3) \, f^*_3(t_4); \nonumber \\
\kappa^{(2)}_p(t) \propto - \lambda^2 \, \int\limits_{t_0}^t dt_3 \, a^{D-1}(t_3) \int\limits_{t_0}^t dt_4 \, a^{D-1}(t_4) \, \int d^{D-1}\vec{q}_1 \, d^{D-1}\vec{q}_2 \, d^{D-1}\vec{q}_3 \times \nonumber \\ \times  \delta^{(D-1)}\Big(\vec{p} + \vec{q}_1 + \vec{q}_2 + \vec{q}_3\Big) \, f_p(t_3) \, f_p(t_4) \, f_1(t_3) \, f^*_1(t_4) \, f_2(t_3) \, f^*_2(t_4) \, f_3(t_3) \, f^*_3(t_4), \label{nkappa}
\end{eqnarray}
where the upper index $2$ indicates that these contributions come from the second loop. (See also \cite{Berges:2004yj} for a similar discussion in the case of the standard modes (\ref{modes}), when $\kappa_p$ is not generated.) The integrations over $t_3$ and $t_4$ come from the two vertexes of the sunset diagram. To clarify the physical meaning of the obtained secular memory effects let us make several comments about the simplest situations. 

First, let us discuss again Minkowski space--time, where $a(t) = 1$ and the natural choice of $f_p(t)$ is such as in eq. (\ref{modes}). Then the integrals over $t_3$ and $t_4$ in (\ref{nkappa}) give, as $t-t_0 \to \infty$, the secular multiplier $(t-t_0)$, which is accompanied by the delta--function establishing the energy conservation, $\delta\left(\omega_p + \omega_1 + \omega_2 + \omega_3\right)$, under the integral over $q_1$, $q_2$ and $q_3$. The argument of the delta--function is never zero.

Thus, for the Poincar\'e invariant state the leading contributions (\ref{nkappa}) do not grow with time: $n_p^{(2)} \sim 0$ and $\kappa_p^{(2)} \sim 0$. This is in accordance with the discussion in the section {\bf 3}. Such terms as (\ref{nkappa}) have been dropped out from (\ref{correc}) at the very beginning on the grounds of the energy--momentum conservation (see e.g. \cite{Berges:2004yj} for the related discussion).

Second, in Minkowski space--time instead of $f_p(t)$ from (\ref{modes}), one in principle could choose e.g. modes of the form 

\begin{equation}
    f_p(t) = \alpha_p \, \frac{e^{-i\, \omega_p \, t}}{\sqrt{2\,\omega_p}} + \beta_p \, \frac{e^{i\, \omega_p \, t}}{\sqrt{2\,\omega_p}},\label{modetil}
\end{equation}
with the condition $\left|\alpha_p\right|^2 - \left|\beta_p\right|^2 = 1$, to have the proper commutation relations for $\hat{a}_{\vec{p}}$ and $\hat{a}^+_{\vec{p}}$. One should also demand that $\beta_p \to 0$, as $\omega_p \to \infty$, to have the proper (Hadamard) behaviour of CF for light--like separations of their points. 

The corresponding Fock space ground state, $\hat{a}_{\vec{p}}\, \left|\alpha,\beta\right\rangle = 0$, is not Poincar\'e invariant, because in such a situation tree--level CF are not functions of only Lorentz invariants: look at e.g. (\ref{Keldysh}) with $f_p(t)$ from (\ref{modetil}) and make the reverse spatial Fourier transformation. In this case the time translational symmetry is broken already at the tree--level: Keldysh propagator depends on its each time argument separately.

Furthermore, from (\ref{nkappa}) with $f_p(t)$ from (\ref{modetil}) it follows that $n_p^{(2)}$ and $\kappa_p^{(2)}$ grow with time as $\lambda^2 \, (t-t_0)$, with non--zero coefficients proportional to powers of $\beta_p$ from (\ref{modetil}) (and $\beta_p^*$). This growth appears due to the interference terms between $\alpha$ and $\beta$ contributions to (\ref{modetil}). Furthermore, the growth of anomalous expectation values $\kappa_p$ signals that the Fock space ground state, $\hat{a}_{\vec{p}}\, \left|\alpha,\beta\right\rangle = 0$, with which we have started at $t_0$, is changing in time. That is drastically different from the situation (\ref{cond}) for the Poincar\'e invariant state. 

At this stage one already can draw a lesson that secular memory effects at every loop order do depend on the choice of a basis of modes, in general, and on the initial state, in particular. In fact, we could have chosen an initial state of the form (\ref{initial}) with $n_p^0 \neq 0$ and $\kappa_p^0 \neq 0$ rather than Fock space ground state, $\hat{a}_{\vec{p}}\, \left|\alpha,\beta\right\rangle = 0$, meanwhile keeping the same basis of modes (\ref{modetil}). Then, the form of the loop corrections in the limit (\ref{limit}) would have been different both from (\ref{nkappa}) and from (\ref{correc1}), (\ref{correc}) \cite{Akhmedov:2011pj}.

On general physical grounds one can predict that the Fock space ground state $\hat{a}_{\vec{p}}\, \left|\alpha,\beta\right\rangle = 0$ will evolve towards the Poincar\'e invariant state $\left|1,0\right\rangle$, if the initial $\beta_p$ is very small. The Poincar\'e invariance, however, will be broken by the presence of the eventual spatially homogeneous particle density of the plankian form. 

In fact, to see the described phenomenon mathematically one can perform the Bogoliubov rotation from (\ref{modetil}) to (\ref{modes}). After such a rotation there will be already some non--zero initial values $n_p^0$ and $\kappa_p^0$ for the proper modes (\ref{modes}). For small $\beta_p$ in (\ref{modetil}) the value of $\kappa^0_p$ after the rotation will be small (proportional to $\beta_p$). In such a case one can apply the linearized approximation in powers of $\kappa_p^0$. Then in the kinetic approximation it can be shown that $\kappa_p(t)$ relaxes to zero. At the same time $n_p(t)$ relaxes to the plankian distribution, for the same reason as it was explained in the section {\bf 5}. 

For large values of $\kappa_p^0$ the problem remains unsolved. (See, however, \cite{Akhmedov:2011pj} for a system of kinetic equations for $n_p$ and $\kappa_p$ simultaneously.) Furthermore, we should not forget about the option of the creation of a condensate, in which $\kappa_p$ remains non--zero even for the proper modes (\ref{modes}), but perhaps with modified dispersion relation. Then, as we know from observational data, the thermalization still can happen, but over a different Poincar\'e invariant state corresponding to a non--zero expectation value of the field. This is not yet a well explored field from the theoretical point of view. Namely we do not yet quite understand the dynamical creation of condensates.

{\bf 7.} Let us continue with the consideration of the theory (\ref{action}) over a generic metric of such a form as (\ref{metric}). Then modes $f_p(t)$ in (\ref{phi}) contain the volume factor, $\left[a(t)\right]^{(D-1)/2}$. This means that such components of the two--point function (\ref{Keldysh}) as $n_p$ and $\kappa_p$ are attributed to the comoving volume, i.e. are always accompanied by the factors of the form $\left[a(t)\right]^{D-1}$ in CF. Hence, all the ``kinetic processes'' that we discuss below are taking place in comoving volume. Then, if the state under consideration is spatially homogeneous the kinetics does not depend on whether spatial sections expand or shrink.

This is a very important and counterintuitive observation. In fact, as we will see below, this observation invalidates the general argument that in expanding universe there is a slow particle production process. Hence, it is generally argued that the density of the created particles is diluted by a rapid expansion of spatial sections. Perhaps that can be true in a space--time of large curvature, although we are not familiar with any calculation from first principles which supports this ``general physical'' argument. But, in any case, in the expanding Universe of GUT scale curvature there are no bound states and one should consider exact modes of fundamental fields rather than point like particles. The exact modes expand together with spatial sections. As a result it is not a surprise that ``kinetic processes'' with their participation should take place in comoving volume.

An example of such a space--time as (\ref{metric}) is the expanding Poincar\'e patch (EPP) of dS space--time, for which $a(t) = e^t$, if the Hubble constant is set to one. In EPP one usually considers so called Bunch--Davies (BD) modes \cite{Bunch:1978yq}:

\begin{equation}
f_p(t) \propto e^{\frac{D-1}{2} \, t} \, H^{(1)}_{i\mu}\left(p \, e^{-t}\right), \quad {\rm where} \quad \mu = \sqrt{m^2 - \left(\frac{D-1}{2}\right)^2}, \label{BD} 
\end{equation}
and $H_\nu^{(1)}(x)$ is the Hankel function of the first kind. The aforementioned volume factor is clearly seen in the modes as the multiplier of the Hankel function. 

Such a choice of modes as (\ref{BD}) is based on the fact that when their physical momentum is much larger than the corresponding Compton value, $p \, e^{-t} \gg |\mu|$, the functions $f_p(t)$ behave as single waves\footnote{Note that for any fixed $p$ this happens at past infinity of EPP, i.e. as $t\to - \infty$.}. As a result the corresponding CF have the proper Hadamard behaviour for light--like separations of their points. 

Furthermore, the tree--level two--point Wightman function, which is calculated for the Fock space ground state $\hat{a}_{\vec{p}} \, \left|BD\right\rangle = 0$ with the modes (\ref{BD}), depends on the isometry invariants \cite{Bunch:1978yq} (see also \cite{Akhmedov:2013vka} for a review): 

\begin{eqnarray}
\Big\langle BD \Big| \phi(\underline{x_1}) \, \phi(\underline{x_2}) \Big| BD\Big\rangle = G_0\Big[Z_{12} + i \, 0 \, {\rm sign} \left(t_1 - t_2\right)\Big], \quad {\rm where} \nonumber \\ Z_{12} = \cosh\Big(t_1 - t_2\Big) -  \frac12 \, e^{t_1+t_2} \, \Big(\vec{x}_1 - \vec{x}_2\Big)^2. \label{Zxy}
\end{eqnarray}
The isometry invariant $Z_{12}$ is the function of the geodesic distance between $\underline{x}_1$ and $\underline{x}_2$. Using the Wightman function one can build the retarded, advanced, Keldysh and, if necessary, the Feynman propagators.

The Wightman function (\ref{Zxy}) is maximally analytic on the cutted complex $Z_{12}$--plane. Moreover, loop corrected CF respect the dS isometry, if the calculation of loop contributions is performed over the invariant BD state and for the initial conditions imposed at the light--like past infinity of EPP (boundary of the patch) \cite{Higuchi:2010xt}, \cite{Hollands:2010pr}, \cite{Akhmedov:2013vka}, \cite{Polyakov:2012uc}. Namely, loop corrected CF are analytic functions of the isometry invariants. Furthermore, in such a case one can perform the analytical continuation to the sphere in all loop integrals \cite{Higuchi:2010xt}, \cite{Hollands:2010pr}, \cite{Gorbenko:2019rza}. The situation in many respects in similar to the less complicated one in anti de Sitter space--time \cite{Akhmedov:2020jsi}.

This is all nice and good, but ``The devil is in the details''. Let us point out that secular effects are still present even in such an isometry symmetric situation. Furthermore, equations (\ref{nkappa}) do contain the secular effect, which would be of interest for us. The problem is that the effect in question, while being secular, is not of memory type: corrections do grow with time $t = (t_1 + t_2)/2$, but there is no any IR divergence. 

In fact, in the symmetric situation the corrected by the two--loop sunset diagram Wightman function has the following form\footnote{See \cite{Krotov:2010ma}, \cite{Jatkar:2011ju} and \cite{Akhmedov:2019cfd} for the similar calculation of the one loop correction in $\phi^3$ theory.}: 

\begin{equation}
G_{0+2}(Z_{12}) \approx \Big[1 + \lambda^2 \, K \, \log Z_{12}\Big] \, G_0(Z_{12}), \quad \left|Z_{12}\right|\to \infty, \label{GZ}    
\end{equation}
where $K$ is some constant. This expression does coincide with the correction, which can be found in the same theory on the sphere, when $Z_{12}$ is just a cosine of the radial distance on the sphere rather than the hyperbolic distance in dS space--time.

At the same time, in dS space--time, taking into account the expression for $Z_{12}$ (\ref{Zxy}), one can see that in the limit $\left|t_1 - t_2\right| \gg (t_1 + t_2)/2$, for large time--like separations, the loop corrected propagator (\ref{GZ}) contains the factor $\log Z_{12} \sim \left|t_1 - t_2\right|$. This is the secular effect of the first kind, which was mentioned in the section {\bf 4}. This effect is not very interesting for us, because it is never of memory type. 

At the same time, in the limit $\left|t_1 - t_2\right| \ll (t_1 + t_2)/2 \equiv t$, for large space--like separations, the loop corrected propagator (\ref{GZ}) contains the factor of $\log Z_{12} \sim t$, which would be the secular memory effect of interest for us. Thus, both secular effects seem to be present and are related to each other via dS isomery and analytical continuation in $Z_{12}$. 

However, the corrected CF (\ref{GZ}) does not depend on the infrared cutoff $t_0$. {\bf Namely for the invariant BD state we have secular growth, $\lambda^2 \, t$, rather than secular IR divergence, $\lambda^2 \, (t-t_0)$.}  It means that in the situation under consideration the initial Cauchy surface can be shifted to the past infinity, as it is the case in equilibrium. 

In all, dS symmetric situation in many respects is similar to the Poincar\'e invariant one: CF depend on distances between their points rather than on the coordinates separately. The resummation of secular effects in such a case is straighforward and was done in many places (see e.g. \cite{Moreau:2018lmz} and \cite{Jatkar:2011ju}). At leading order the result is a (complex) shift of mass.

But let us see in grater detail how does it happen that on the one hand there is a secular effect following from (\ref{nkappa}), as it seem to happen only in non--stationary situations. While, on the other hand, there is no any IR divergence, which should be cutted by $t_0$, as it is the case in equilibrium. 

{\bf 8.} Consider for simplicity the massive scalar field $m > (D-1)/2$. The point is that the modes (\ref{BD}) behave as

\begin{equation}
    f_p(t) \sim e^{\frac{D-1}{2} \, t} \, \Big[C_+ \, e^{i\,\mu \, t} + C_- \, e^{-i\, \mu \, t}\Big], \quad \mu = \sqrt{m^2 - \left(\frac{D-1}{2}\right)^2} \label{modeasym}
\end{equation}
at future infinity --- when the physical wave--length exceeds the Compton's one, $p e^{-t} \ll |\mu|$. Thus, when $m > (D-1)/2$ the modes oscillate at future infinity rather than homogeneously decay, as is the case when $m < (D-1)/2$. As a result, the $m > (D-1)/2$ case resembles the Minkowski space--time kinetic situation \cite{Akhmedov:2017ooy}, unlike the fields with $m < (D-1)/2$. 

Moreover, as is shown in \cite{Akhmedov:2017ooy}, for small enough mass below the threshold $m = (D-1)/2$, multiple point CF also contain secular effects. And lower is mass, higher point CF grow as all their points are taken to future infinity. In our opinion this fact complicates drastically the resummation of the leading corrections at least in non invariant situations, when the BD state is perturbed. So to simplify the discussion we restrict our attention to $m > (D-1)/2$ case\footnote{Even for such massive fields we will see that there can be strong secular memory effects after loop resumation in non--invariant situations. It goes without saying that the same effects are stronger for the light fields.}.

Thus, for $m > (D-1)/2$ case from (\ref{nkappa}) we obtain for the modes (\ref{BD}) with low physical momentum, $p \, e^{-t} \ll |\mu|$, that \cite{Akhmedov:2013vka}:

\begin{equation}
n_p^{(2)}(t) \propto \lambda^2 \log \left(\frac{\mu}{p\, e^{-t}}\right) \sim \lambda^2 \, t, \quad \text{and}\quad \kappa_p^{(2)}(t) \propto \lambda^2 \log \left(\frac{\mu}{p\, e^{-t}}\right) \sim  \lambda^2 \, t, \label{nkappa1}
\end{equation}
as $t\equiv (t_1 + t_2)/2 \to \infty$. These expressions agree with (\ref{GZ}) \cite{Krotov:2010ma}, \cite{Akhmedov:2019cfd}. The secular growth in (\ref{nkappa}), leading to (\ref{nkappa1}), appears from the interference terms between $C_\pm$ contributions to $f_p(t_{3,4})$ in (\ref{modeasym}). 

At the same time the dependence on $t_0$ does not appear in (\ref{nkappa1}) for the following reason. The point is that at past infinity every mode (\ref{BD}) experiences the blue shift, $p e^{-t} \gg |\mu|$, and behaves as a plane wave in Minkowski space--time. Namely that is because the past infinity limit in EPP coincides with the UV one for the physical momentum $p e^{-t}$. Hence, if one takes $t_0 \to - \infty$ for the BD modes (\ref{BD}) the time period between $k \, e^{-t_0} \ll k \, e^{-t} \ll \mu$ for fixed $k = p, \, q_{1,2,3}$ in (\ref{nkappa}) does not lead to any secular loop contribution. 

In other words, the situation for the BD modes whose physical wavelength did not exceed their Compton one, $p \, e ^{-t} \gg |\mu|$, is essentially the same as for plane waves over the Poincar\'e invariant state in Minkowski space--time. That is the reason why the growth (\ref{nkappa1}) is cutted from below by $\log \mu$ rather than by $\log\left(p \, e^{-t_0}\right)$.

Consider, however, a different state in EPP: e.g. fix a Cauchy surface at $t_0 > - \infty$ and take on it an initial Hadamard state of the form (\ref{initial}) with generic $\kappa^0_p$ and $n_p^0$ for the BD modes (\ref{BD}). In such a case there are finite level populations and anomalous averages per physical volume. Then, one cannot take $t_0 \to - \infty$, because otherwise the level density per physical volume will become infinite. For the state in question the dS isometry is violated at tree--level: tree--level Wightman propagator will not be just a function of the invariants.

What is important for our discussion is that in the situation under consideration the dependence on $t_0$ will appear at every loop order: there will be the secular divergence similar to the one in flat space--time, which was discussed in the sections before {\bf 7}. 

If there will be enough time for such a state to thermalize during the time period $p \, e^{-t_0} \ll p \, e^{-t} \ll \mu$, the situation will not be much different from the dS symmetric one. Then, in first place, the dependence on $t_0$ in CF will disappear before $p \, e^{-t} \sim \mu$. In such a case there also will be a clear grading between diagrams: One can show \cite{Akhmedov:2019cfd} that the leading secular contributions will come only from the diagrams containing chains of sunset bubbles. While corrections into the internal legs of the loops (inside sunset bubbles) will be suppressed by higher powers of $\lambda^2$, which are not accompanied by higher powers of the secular factor $t-t_0$. 

As a result, in the situation in question to resum the leading corrections one has to solve a linear Dyson--Schwinger equation rather than a non--linear integro--differential equation of the kinetic type\footnote{We would like to thank A.Polyakov and F.Popov for pointing this important issue to us.}. 

Otherwise, if there is no enough time for the thermalization of the state with $\kappa^0_p$ and $n_p^0$ in the regime $p \, e^{-t_0} \ll p \, e^{-t} \ll \mu$, then the problem of loop resummation for generic initial states remains to be unsolved. Moreover, in space--times in which a rapid expansion is sandwiched between flat spaces (consider e.g. metrics from \cite{Akhmedov:2019cfd}, \cite{Birrell:1982ix} or \cite{Akhmedov:2017dih}), the leading secular loop corrections also come from the loops in internal legs \cite{Akhmedov:2019cfd}. Then, the problem of resummation of loop corrections for generic initial states also remains unsolved. 

{\bf 9.} Furthermore, consider contracting Poincar\'e patch (CPP) of dS space--time. It is described by the metric (\ref{metric}) with $a(t) = e^{-t}$ and is just time--reversal of the EPP. We consider this patch because it is instructive to do that for the integrity and because to understand the situation in global dS space--time one has to consider CPP together with EPP. In fact, the global dS contains both patches and will be discussed below.
 
For the BD Fock space ground state, corresponding to the modes (\ref{BD}), the tree--level Wightman function in CPP is the same as in EPP (\ref{Zxy}). Furthermore, for spatially homogeneous states the situation in CPP is just time reversal of the situation in EPP. Namely, expressions (\ref{Keldysh}) and (\ref{nkappa}) are also valid in CPP. 

However, as a result of this time reversal, the calculation of (\ref{nkappa}) for the BD modes (\ref{BD}) in CPP leads to a physically distinct answer from the EPP one. {\bf Namely, exactly for the same reason as we have the secular growth (\ref{nkappa1}) in EPP, in CPP there is the secular divergence \cite{Akhmedov:2013vka}}:

\begin{eqnarray}
    n_p^{(2)}(t) \quad {\rm and} \quad \kappa_p^{(2)}(t) \propto 
    \left\{
  \begin{array}{cc}
    \lambda^2 \log \left(\frac{p \, e^{t}}{p\, e^{t_0}}\right) \sim \lambda^2 \, (t - t_0)   &  \quad {\rm for} \quad p \, e^{t} < \mu, \\
     \lambda^2 \log \left(\frac{\mu}{p\, e^{t_0}}\right) &  \quad {\rm for} \quad p \, e^{t} > \mu.\\
  \end{array}
\right.
\end{eqnarray}
Thus, even though the initial state is dS isometric the symmetry is broken by loop corrections: loop corrected CF are not functions of the invariants anymore. They also depend on the position of the initial Cauchy surface $t_0$. Such a situation cannot be described by the analytical continuation from the sphere. It goes without saying that for more generic initial states loop corrections will also contain IR divergences in CPP. 

But it is interesting that for spatially homogeneous initial states in CPP with small deviations from the BD state the symmetry is restored after the resummation of the leading loop corrections  \cite{Akhmedov:2013vka}. At the same time, for the strong deviations from the BD state even for the spatially homogeneous states there is a blow up of the comoving particle density during a finite proper time \cite{Akhmedov:2013vka}. This means that backreaction cannot be neglected in such a situation.


{\bf 10.} Let us conclude our discussion with some comments about the situation in global dS space--time.
There are different types of metrics that cover the global dS space--time. One type is as follows:

\begin{equation}
    ds^2 = \frac{1}{\eta^2} \, \left[- d\eta^2 + d\vec{x}^2\right], \quad \eta \in (-\infty, +\infty).
\end{equation}
For $\eta = - e^{-t} \in (-\infty, 0)$ this metric covers the EPP and for $\eta = e^{-t} \in (0, +\infty)$ it covers the CPP. In this case the Cauchy surfaces $\eta = $const are not compact and flat. 

Then loop corrections in global dS contain those in CPP and EPP simultaneously. As a result we obviously encounter IR secular memory effect in global dS for any initial state, including the so called Euclidian one \cite{Krotov:2010ma}, \cite{Akhmedov:2013vka}, in which the Wightman function has the same form as for the BD state in EPP or CPP. The the situation in global dS is in many respects similar to the one in CPP \cite{Akhmedov:2013vka}, \cite{Akhmedov:2013xka}.

Another metric covering global dS is:

\begin{equation}
    ds^2 = - d\tau^2 + \cosh^2 \tau \, d\Omega^2,
\end{equation}
where $d\Omega^2$ is the line element of the unit $(D-1)$--dimensional sphere. In this case Cauchy surfaces are compact spheres. Loop corrections can be calculated also in such a Cauchy slicing of the global dS \cite{Akhmedov:2019cfd}, \cite{Akhmedov:2012dn}. The result is the same as for the non--compact slicing mentioned above. Here one also encounters IR secular memory effect. The dS isometry is violated by loop corrections.

A fair question at this point is as follows: {\bf How can it be that loop corrections in global dS metric are different from those in EPP metric, if we restrict our selves to the same patch?} In fact, there is just a coordinate transformation which should relate any physical quantity in both coordinate systems. 

However, the essential ingredient of the question under consideration is the restriction to the same patch. {\bf The point is that secular memory effects are sensitive to the initial and boundary conditions, due to their IR memory nature.} Because of that, as we have mentioned at the very beginning of this note, to unambiguously define the behaviour of the loop corrections (the time evolution of CF) one has to specify the initial Cauchy surface (its geometry), a basis of modes and a state build with the use of the corresponding creation and annihilation operators. 

Finally let us mention the situation in static patch of dS space--time. In \cite{Akhmedov:2020ryq}, \cite{Anempodistov:2020oki} it is shown that in this patch states with plankian distribution (for exact modes) of generic temperature do not possess the standard properties of the thermal states in flat space--time. E.g, if the temperature is not equal to the Gibbons--Hawking one, the backreaction on the background geometry is not negligible in the sense similar to the one described in \cite{Ho:2018jkm}. Namely the stress--energy tensor calculated with the use of the corresponding tree--level two point function blows up at the horizon. This rises the question if it is possible to put a gas with generic temperature in dS space--time of high curvature. Furthermore, in \cite{Popov:2017xut} it is shown that even the state with the canonical temperature in dS space--time does not possess some of the standard properties of thermal states in flat space--time. We mention this just to stress again that the intuition gained by consideration of flat space QFT is not directly applicable to the curved space--time.

{\bf 11.} In all, we can rephrase the question formulated in the main body of this note. In Minkowski space--time the thermal states over the Poincar\'e invariant vacuum seem to be attractors of the unitary evolution for a reasonable initial conditions. The question is if there are any attractor states in expanding space--times and for what initial conditions these states are attractors.

Here we tried to convince the reader that the answer on this question is not so obvious and the intuition gained by the consideration of flat space stationary situations is not of any use in curved space--time.


I would like to acknowledge valuable discussions of these issues with P.Anempodistov, K.Basarov,  D.Diakonov, O.Diatlyk, H.Godazgar, K. Kazarnovsky, V.Losyakov, J.Maldacena, D.Marolf, I.Morrison, A.Morozov, U.Moschella, E.Mottola, F. Popov, A.Polyakov, T.Prokopec, A.Roura, V.Rubakov, A.Semenov, J.Serreau, A,Starobinsky,  D.Trunin, S.Theisen, G.Volovik, R.Woodard. The work was supported by the grant from the Foundation for the Advancement of Theoretical Physics and Mathematics ``BASIS'', by RFBR grant 19-02-00815 and by Russian Ministry of education and science.

\end{document}